\PassOptionsToPackage{table}{xcolor}
\documentclass[10pt,conference]{IEEEtran} 

\usepackage[utf8]{inputenc}
\usepackage[hidelinks]{hyperref}
\usepackage{booktabs}
\usepackage{tcolorbox}
\usepackage{amsmath}
\usepackage{amssymb}
\usepackage{nccmath}
\usepackage{paralist}
\usepackage[shortlabels]{enumitem}
\usepackage{tabularx}
\usepackage{graphicx}
\usepackage{multicol,tabularx}
\usepackage{color, colortbl}
\usepackage{array}
\usepackage{makecell}
\usepackage{multirow}
\usepackage{csvsimple}
\usepackage{xcolor}
\usepackage{xspace}
\usepackage{colortbl}
\usepackage{siunitx}
\usepackage[framemethod=tikz]{mdframed}
\usepackage{balance} %

\usepackage{listings}
\newcommand{\lsin}[1]{\lstinline[columns=fullflexible,keepspaces=true,language=Java,basicstyle=\footnotesize\ttfamily]{#1}}

\lstdefinelanguage{Java}{
  columns=fullflexible,keepspaces=true,basicstyle=\scriptsize\ttfamily,
  numbers=left,
  xleftmargin=5.0ex,
  escapechar=@,
  sensitive=true,
  otherkeywords={},
  morekeywords=[1]{int,while,if,public,static,else,return,new,for,return,void},
  keywordstyle={[1]\bfseries\color{blue}},
  numberstyle=\tiny\color{black},
  rulecolor=\color{black},
}

\newsavebox{\mylisting}

\definecolor{lightgray}{gray}{0.9}
\definecolor{Gray}{gray}{0.9}
\definecolor{lightcyan}{rgb}{0.88,1,1}

\mdfdefinestyle{mystyle}{%
  rightline=false,
  innerleftmargin=5,
  innerrightmargin=5,
  outerlinewidth=1pt,
  topline=false,
  leftline=true,
  bottomline=false,
  skipabove=\topsep,
  skipbelow=0pt,
  backgroundcolor = lightgray,
}

\newcommand{\head}[1]{\par\noindent\textbf{#1:}\space}

\newcommand{\SLoC}{\emph{SLoC}\xspace}
\newcommand{\TSS}{\emph{TSS}\xspace}
\newcommand{\BR}{\emph{BR}\xspace}
\newcommand{\RT}{\emph{RT}\xspace}    %
\newcommand{\NPC}{\emph{NPC}\xspace}
\newcommand{\NTE}{\emph{NTE}\xspace}

\newcommand{\PP}{\ensuremath{\mathcal{P}}\xspace}         %
\newcommand{\PR}{\ensuremath{\mathcal{P}_{S}}\xspace}
\newcommand{\PS}{\ensuremath{\mathcal{P}_{S}}\xspace}
\newcommand{\T}{\ensuremath{\mathcal{T}}\xspace}
\newcommand{\TT}{\ensuremath{\mathcal{T}}\xspace}         %
\newcommand{\TR}{\ensuremath{\mathcal{T}_{S}}\xspace}
\newcommand{\TS}{\ensuremath{\mathcal{T}_{S}}\xspace}
\newcommand{\SL}{\ensuremath{\mathcal{L}}\xspace}         %
\newcommand{\SLR}{\ensuremath{\mathcal{L}_{S}}\xspace}    
\newcommand{\SLS}{\ensuremath{\mathcal{L}_{S}}\xspace}    
\newcommand{\SLP}{\ensuremath{\mathcal{L}_{P}}\xspace}

\newcommand{\Cooo}{$\mathcal{C}_{\PP\T\SL}$\xspace}
\newcommand{\Coor}{$\mathcal{C}_{\PP\T\SLR}$\xspace}
\newcommand{\Coso}{$\mathcal{C}_{\PP\TS\SL}$\xspace}
\newcommand{\Coop}{$\mathcal{C}_{\PP\T\SLP}$\xspace}
\newcommand{\Cosr}{$\mathcal{C}_{\PP\TS\SLR}$\xspace}
\newcommand{\Cosp}{$\mathcal{C}_{\PP\TS\SLP}$\xspace}
\newcommand{\Cssr}{$\mathcal{C}_{\PS\TS\SLR}$\xspace}
\newcommand{\Cssp}{$\mathcal{C}_{\PS\TS\SLP}$\xspace}
\newcommand{\Csoo}{\xspace{\color{red}{$\mathcal{C}_{\PS\T\SL}$}\xspace}}
\newcommand{\Csor}{\xspace{\color{red}{$\mathcal{C}_{\PS\T\SLR}$}\xspace}}
\newcommand{\Csso}{\xspace{\color{red}{$\mathcal{C}_{\PS\TS\SL}$}\xspace}}
\newcommand{\Csop}{\xspace{\color{red}{$\mathcal{C}_{\PS\T\SLP}$}\xspace}}

\newcommand{\sfs}[1]{\textsf{\small #1}}   %

\usepackage[backend=biber, abbreviate=true, dateabbrev=true, alldates=year, isbn=false, doi=true, urldate=comp, url=false, maxbibnames=8, backref=false, style=numeric-comp, language=american, giveninits=true, sortcites=true, sorting=none]{biblatex}
\AtEveryBibitem{\clearlist{location}} %
\AtEveryBibitem{\clearfield{series}} %
\AtEveryBibitem{\clearlist{language}} %

\usepackage{citation-cleaner}
\usepackage{adjustbox}

\title{The Impact of Program Reduction \\ on Automated Program Repair}

\author{
\IEEEauthorblockN{Linas Vidziunas}
\IEEEauthorblockA{\textit{Simula Research Laboratory} \\
Oslo, Norway \\
linasvidz@simula.no}
\and
\IEEEauthorblockN{David Binkley}
\IEEEauthorblockA{\textit{Loyola University Maryland} \\ 
Baltimore, MD, USA\\
binkley@cs.loyola.edu}
\and
\IEEEauthorblockN{Leon Moonen}
\IEEEauthorblockA{\textit{Simula Research Laboratory} \& \\
\textit{BI Norwegian Business School} \\
Oslo, Norway \\
leon.moonen@computer.org}
}

\makeatletter
\def\ps@IEEEtitlepagestyle{%
  \def\@oddfoot{\mycopyrightnotice}%
  \def\@evenfoot{}%
}
\def\mycopyrightnotice{%
  \hspace*{3mm}\includegraphics[width=2cm]{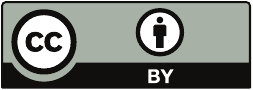}%
  \hspace*{2mm}\raisebox{2.5mm}{%
   	  \parbox{\columnwidth}{\footnotesize This work is licensed under a Creative Commons \\ Attribution 4.0 International (CC BY 4.0) license.}%
   	  \hspace*{-68pt}\mbox{1}\hspace{20pt}\fbox{\parbox{.86\columnwidth}{\footnotesize\textsl{Accepted for publication in the 40th IEEE International Conference on Software Maintenance and Evolution (ICSME 2024).}}}%
  }%
  \gdef\mycopyrightnotice{}%
}
\makeatother

\begin{document}

\maketitle

\noindent \begin{abstract}

Correcting bugs using modern Automated Program Repair (APR) can be both
time-consuming and resource-expensive.
We describe a program repair approach that aims to improve the scalability 
of modern APR tools.
The approach leverages program reduction in the form of program slicing 
to eliminate code irrelevant to fixing the bug, 
which improves the APR tool's overall performance.
We investigate slicing's impact on all three phases of the repair process:
fault localization, patch generation, and patch validation. 
Our empirical exploration finds that the proposed approach on average 
enhances the repair ability of the TBar APR tool, 
but we also discovered a few cases where it was less successful. 
Specifically, on examples from the widely used Defects4J dataset, 
we obtain a substantial reduction in median repair time, 
which falls from 80 minutes to just under 18 minutes. 
We conclude that program reduction \emph{can}  
improve the performance of APR without degrading repair quality, 
but this improvement is not universal.

\end{abstract}

\begin{IEEEkeywords}
automated program repair,
dynamic program slicing,
fault localization,
test-suite reduction,
hybrid techniques.
\end{IEEEkeywords}

\section{Introduction}

\noindent 
Automated program repair (APR) is an active research area with growing tool
support~\cite{monperrus2018:automatic,legoues2019:automated}.
However, APR struggles with scalability as it grows resource-intensive and
time-consuming for larger programs.  
Of its three phases, fault localization, patch generation, and patch
validation, this struggle is more pronounced in the first and third phases.
For example, poor performance in the first phase can lead to poor overall
performance because the actual faulty statement is found far down the
list of suspicious statements~\cite{liu2020:efficiency,liu2019:you}, 
leading to what has been termed the ``patch explosion'' problem.

This paper aims to study the improvement possible using program reduction,
specifically, that made possible through program slicing.
Fixing a reduced program instead of the original program is expected to provide
benefits to all three phases.
For example, fault localization should be more focused when working with the
reduced program.
The impact on patch generation is less direct.
The reduced program is, de facto, more cohesive, than the original program;
thus, sampling potential patches from the reduced program should provide a higher
probability of success. 
Finally, running the simpler reduced program should lower patch validation
time.  
This is especially true because we introduce a test-suite reduction technique
based on the reduced program.
As others have noted, for test-based APR, the cost of patch validation depends
heavily on the size of the test suite, while at the same time, much of a test
suite is irrelevant to fixing a given bug~\cite{legoues2012:genprog,legoues2019:automated}.

To demonstrate the effectiveness of combining program reduction and APR we use
TBar~\cite{liu2019:tbar}, a state-of-the-art template-based APR tool, and select
bugs from the widely studied Defects4J dataset~\cite{just2014:defects4j}.
We choose template-based APR because it is one of the most effective approaches
for generating patches~\cite{liu2020:efficiency}.
We use Defects4J to illustrate the effectiveness of our approach on several
real applications and to facilitate comparison with previous
work~\cite{just2014:defects4j}.

\head{Significance and Impact}
Empirical evidence~\cite{weimer2013:leveraging,legoues2013:current,le2016:history} has shown that finding a
valid patch often requires exploring a significant proportion of the patch
space, which can grow quite expensive.
We aim to show how a reduced program, such as that produced by program slicing,
can \emph{effectively} identify and eliminate code and tests that
are irrelevant to the patching of a bug without adversely affecting the quality
of the repair.
Thus, in essence, we aim to improve program repair by reducing the size of the
program being repaired.
Two things are essential to this process.
First, the reduced program must replicate the erroneous behavior of the bug
being repaired, and, second, a patch for the reduced program must also be a
patch for the original program.
Satisfying these two requirements expands the set of programs to which modern
APR tools can be applied.

\head{Contributions}
This paper makes the following contributions:
\begin{compactitem}\renewcommand{\labelitemi}{$\star$}
\item We empirically investigate the impact of program reduction on the three phases of automated program repair: fault localization, patch generation, and patch validation. 
\item We discuss a dynamic procedure for test-suite reduction that is based on program reduction and introduce a new metric \NTE (Number of Tests Executed) to study the impact that program reduction has on patch validation.
\item We consider eight configurations that involve combinations of a reduced program, a reduced test suite, and a reduced suspicious-locations list.
\item Our results show a reduction in TBar's runtime without degrading its repair quality. 
On examples from the widely studied Defects4J dataset, we obtain substantial reductions that approach an order of magnitude. 
\item To support open science, %
we make a replication package available via Zenodo.\footnote{%
~\url{https://doi.org/10.5281/zenodo.13074333}
}
\end{compactitem}

\section{Motivating Example}

\noindent 
Consider the repair of Defects4J~\cite{just2014:defects4j} bug
Lang-10 using the TBar APR tool~\cite{liu2019:tbar}.
It takes TBar eight hours to find a plausible patch for this bug.
During this time, it examines 1938 incorrect patches.
The long processing time can be explained by the following observations.
First, TBar spends significant time mutating suspicious statements that are not
the actual faulty statement: fault location returns a ranked list of 261
suspicious statements, on which the faulty location has Rank 82. 
Second, the test suite for this bug includes two failing tests and 2196 passing
tests, most of which are irrelevant to fixing the bug; thus, they can be
omitted without adversely affecting the quality of the repair while saving 
considerable time.
Applying our technique the faulty location improves to Rank 64. 
Furthermore, test-suite reduction reduces the test suite to 73 passing tests,
greatly reducing the patch validation expense.
As a result, the same patch is produced in approximately forty minutes, 
a 91\% reduction in patch time!

\section{Background}

\noindent
This section provides background on 
  automated program repair (APR), 
  the APR tool used (TBar),
  program slicing, and 
  ORBS, the observation-based slicer used in the experiments.

APR needs a specification of correct program behavior to determine if a
generated patch fixes the target bug. 
In the absence of a complete or formal specification, existing APR tools use
the test suite as an oracle for determining the desired behavior.
Thus, testing is the primary method used to ensure the correctness of a
generated patch, causing the cost of patch validation to depend heavily on the
size of the test suite.
As a program evolves its test suite tends to grow in size, which increases the
time and resources required. %

APR commonly involves three-phases: 
(1) \emph{fault localization} produces a ranked list of suspicious locations; 
(2) \emph{patch generation} involves a range of program mutations; 
and (3) \emph{patch validation} checks whether the mutated program passes the
test suite of the program.
In greater detail, APR's first phase employs automated fault localization
to produce a ranked list of suspicious statements based on the bug
being repaired.
The significant expense in this first phase is the quality of the fault
location.
Because APR tools attempt to fix a bug by repeatedly applying a set of
potential patches to each potentially buggy location in the ranked list, the
sooner the actual location of the bug is considered, the better.

The second phase, patch generation, explores the latent patch space that is
typically defined by the set of program mutation operators employed and the
list of suspicious statements.
This phase requires the most innovation from an APR tool.
Typically, this phase makes use of the code itself to derive potential
patches.
This means that the more cohesive the code, the more likely a patch will be
found in the code itself.
As an example, consider an application that maintains inventory, orders, and
taxes.
Code taken from a reduced program that just computes the taxes is more likely
to be relevant to fixing a tax-related bug.

The final step, patch validation, ensures the correct behavior of the repaired
program.
As mentioned previously, the test suite is typically used to ensure that the 
bug is fixed and that the patch does not introduce any unwanted behavior.
In practice, most tests are irrelevant to the bug being repaired and thus a huge
saving can be obtained by avoiding their repetitive execution.
At the bare minimum, the test suite must include at least one failing test that
exposes the bug.  
Passing this test implies the bug has been fixed.
The existence of additional passing tests can constrain the repair when used as 
regression tests.\footnote{~Observe that passing tests are also needed to pinpoint suspicious statements in the initial fault localization step.}
However, such tests provide diminishing returns, 
and as their number grows performance decreases without any improvement in patch quality.
Our test-suite reduction process aims to omit tests that have no impact on patch quality and
thus only affect efficiency.

Template-based APR, which utilizes predefined fix templates to fix specific bugs, 
is widely used in the APR literature, 
and known to be effective at generating correct patches~\cite{durieux2017:dynamic,hua2018:practical,jiang2018:shaping,koyuncu2020:fixminer,le2016:history,liu2019:avatar,liu2018:mining,liu2019:tbar,long2017:automatic,saha2017:elixir,wen2018:contextaware,xin2017:leveraging}.
The TBar tool~\cite{liu2019:tbar} used in this work combines fifteen commonly used fix templates from the APR literature, 
making it one of the most effective APR tools available and achieving a good repair rate on Defects4J.
However, like most APR tools, TBar is highly sensitive to fault localization accuracy and to the size of the test suite~\cite{liu2020:efficiency}. 

The following description of Observation-Based Slicing (ORBS) is taken from
existing work~\cite{binkley2014:orbs, islam2016:porbs, stievenart2023:empirical}.
Readers familiar with ORBS need only skim this section as a reminder.
The key requirement for constructing the reduced program is a sound program
reduction technique.
This enables the repair process to be conducted using the reduced program while
being able to conclude that the repair is valid for the original program.
The program reduction technique we study in this paper is a form of dynamic
program slicing~\cite{korel1988:dynamic}.
Informally, a slice is a subset of a program that preserves the behavior of the
program for a specific \emph{slicing criterion}~\cite{weiser1982:programmers}.
While static slices~\cite{weiser1982:programmers} preserve this behavior \emph{for
all possible inputs}, dynamic slicing~\cite{korel1988:dynamic} does so only for a
selected set of inputs.

The slicing technique used to produce the reduced program studied in our
experiments is observation-based slicing (ORBS), which works by \emph{deleting}
statements and then \emph{observing} the behavior at the slicing criterion.
ORBS takes as input a program \PP, a slicing criterion identified by a
program variable $\nu$, a program location $l$, and a set of inputs
$\mathcal{I}$.
The resulting slice preserves the values of $\nu$ at $l$ for the inputs 
of $\mathcal{I}$.

Operationally, ORBS \emph{observes} behavior by instrumenting the 
program with a side-effect free line that tracks the value of variable
$\nu$ immediately before line $l$.
Then, starting with \PP, ORBS repeatedly forms candidate slices by 
deleting a sequence of one to $\delta$ lines from the current slice.
The candidate is rejected if it fails to compile or produces different 
values for $\nu$.
Otherwise, the candidate becomes the current slice.
ORBS systematically forms candidates until no more lines can be deleted.

While there are more time-efficient dynamic slicers, ORBS is sound and very
accurate.
Its slices are nearly minimal~\cite{binkley2014:orbs}.
Thus, our use of ORBS can be seen as paralleling APR studies that assume
perfect fault localization~\cite{liu2019:tbar,jiang2021:cure,ye2022:neural,meng2023:templatebased,jiang2023:knod,xia2023:automated}.
Faster dynamic slicers, such as JavaSlicer~\cite{wang2020:type} and Slicer4J~\cite{ahmed2021:slicer4j},
and dynamic/static hybrid slicers such as QSES~\cite{stievenart2021:qses}
compute less accurate slices, but can drop slicing time from days to a matter of
minutes.

\section{Experimental Design}

\noindent
This section lays out the experimental design.
It first introduces the bugs we consider and then the metrics we used in the
analysis.
We then describe the steps of the algorithm that we study and the execution
environment used.
Finally, we present the research questions used to evaluate our approach.

\subsection{Subject Bugs Studied}
\noindent
Defects4j is a collection of reproducible bugs widely used in software
engineering research~\cite{just2014:defects4j}.
Our experiments employ Defects4J v2.0.1~\cite{just2014:defects4j}, which has been 
used in prior APR experiments~\cite{jiang2018:shaping,liu2019:tbar}. 
We consider two of the Defects4j systems, \sfs{Lang} and \sfs{Math}, because each
includes a large number of bugs.
\sfs{Lang} provides a collection of utility methods that work with
\sfs{java.lang} including basic numerical methods, concurrency support, as well
as utility methods to help construct methods such as \sfs{hashCode},
\sfs{equals}, etc. 
\sfs{Math} provides common mathematical and statistical methods.
The authors work to limit ``external'' dependencies, which, among other things, makes it easier for
us to experiment with their code. 

\rowcolors{1}{white}{lightcyan}    %
\begin{table}
\vspace*{1ex}
\centering
\adjustbox{max width=\linewidth}{
\begin{tabular}{ll|rrr|rr|r}
  &  
& \multicolumn{2}{c}{\SLoC} & 
\multicolumn{1}{c|}{Slice}           & 
\multicolumn{2}{c}{Passing Tests} & 
\multicolumn{1}{c}{Failing} \\

\multicolumn{2}{c|}{BID}    & 
\multicolumn{1}{c}{Orig}    & 
\multicolumn{1}{c}{Slice}  & 
\multicolumn{1}{c|}{as \%} & 
\multicolumn{1}{c}{Orig}    & 
\multicolumn{1}{c|}{Slice} & 
\multicolumn{1}{c}{Tests} \\

\midrule
Lang   & 10  & 20442                                  & 836   & 4.1  & 2196 & 73   & 2  \\
Lang   & 22* & 18870                                  & 571   & 3.0  & 1823 & 35   & 2  \\
Lang   & 33  & 17302                                  & 907   & 5.2  & 1669 & 67   & 1  \\
Lang   & 39  & 16983                                  & 1437  & 8.5  & 1617 & 93   & 1  \\
Lang   & 43  & 18817                                  & 409   & 2.2  & 1807 & 30   & 1  \\
Lang   & 44  & 18814                                  & 661   & 3.5  & 1784 & 46   & 1  \\
Lang   & 45  & 18777                                  & 549   & 2.9  & 1782 & 38   & 1  \\
Lang   & 47  & 18490                                  & 998   & 5.4  & 1733 & 98   & 2  \\
Lang   & 51  & 17028                                  & 601   & 3.5  & 1630 & 70   & 1  \\
Lang   & 58  & 16961                                  & 1098  & 6.5  & 1594 & 81   & 1  \\
Lang   & 59  & 16943                                  & 571   & 3.4  & 1592 & 52   & 1  \\
Lang   & 63  & 16791                                  & 1146  & 6.8  & 1576 & 36   & 1  \\
Lang   & ~~7   & 21495                                  & 836   & 3.9  & 2260 & 93   & 1  \\
Math   & ~~3*  & 83115                                  & 6048  & 7.3  & 4281 & 81   & 1  \\
Math   & ~~4   & 82856                                  & 5971  & 7.2  & 4235 & 62   & 2  \\
Math   & ~~5   & 81560                                  & 10386 & 12.7 & 4180 & 155  & 1  \\
Math   & ~~6   & 81309                                  & 14883 & 18.3 & 4146 & 146  & 28 \\
Math   & 22  & 68184                                  & 10616 & 15.6 & 3401 & 87   & 2  \\
Math   & 28  & 65079                                  & 3559  & 5.5  & 3265 & 44   & 1  \\
Math   & 32  & 63767                                  & 5403  & 8.5  & 2881 & 53   & 1  \\
Math   & 33  & 63136                                  & 3384  & 5.4  & 2863 & 41   & 1  \\
Math   & 35  & 63064                                  & 2632  & 4.2  & 2838 & 16   & 4  \\
Math   & 50  & 54718                                  & 3335  & 6.1  & 2486 & 29   & 1  \\
Math   & 57  & 45519                                  & 2369  & 5.2  & 2214 & 19   & 1  \\
Math   & 58* & 45271                                  & 3367  & 7.4  & 2185 & 28   & 1  \\
Math   & 60* & 44036                                  & 3234  & 7.3  & 2069 & 44   & 1  \\
Math   & 62  & 47371                                  & 3902  & 8.2  & 2216 & 19   & 1  \\
Math   & 63  & 44354                                  & 4201  & 9.5  & 2148 & 84   & 1  \\
Math   & 65  & 42470                                  & 3573  & 8.4  & 2139 & 55   & 1  \\
Math   & 70  & 41186                                  & 3008  & 7.3  & 2088 & 45   & 1  \\
Math   & 73* & 39234                                  & 3113  & 7.9  & 2047 & 40   & 1  \\
Math   & 75  & 39884                                  & 2936  & 7.4  & 2039 & 39   & 1  \\
Math   & 79  & 39306                                  & 2312  & 5.9  & 2012 & 22   & 1  \\
Math   & 82  & 38672                                  & 2754  & 7.1  & 1966 & 33   & 1  \\
Math   & 85  & 38458                                  & 2760  & 7.2  & 1903 & 40   & 1  \\
Math   & 89  & 30903                                  & 1002  & 3.2  & 1611 & 8    & 1  \\
Math   & 95  & 19934                                  & 936   & 4.7  & 1223 & 22   & 1  \\
Math   & 96  & 19511                                  & 653   & 3.3  & 1195 & 97   & 1  \\
Math   & 98* & 16725                                  & 1389  & 8.3  & 1017 & 62   & 2  \\
\midrule
Lang   & 57  & \multicolumn{6}{c}{Test Source fails to compile}                         \\
Math   & ~~8   & \multicolumn{6}{c}{Fault Localization times out }                          \\
Math   & 30  & \multicolumn{6}{c}{Fault Localization times out }                          \\
Math   & 34  & \multicolumn{6}{c}{Test Source fails to compile}                         \\
Math   & 49  & \multicolumn{6}{c}{Test Source fails to compile}                         \\
Math   & 59  & \multicolumn{6}{c}{Test Source fails to compile}                         \\
Math   & 77  & \multicolumn{6}{c}{Test Source fails to compile}                         \\
Math   & 80  & \multicolumn{6}{c}{Test Source fails to compile}                         \\
\end{tabular}}
\vspace{2ex}
\caption{The subject bugs studied.}
\label{tab:subjects}
\vspace*{-5ex}
\end{table}

Table~\ref{tab:subjects} shows demographics for the 47 subject bugs studied.
Each bug includes a \emph{Bug ID}, BID, its size in non-comment, non-blank
lines of code, \SLoC, and the number of passing and failing tests.
Program size is given for both the original production code (i.e., excluding
the test code) and the slice along with the slice's size as a percent of the
original code.
Finally, in addition to the number of passing tests, the number that apply to
the slice is also included.

The initial collection of subjects included the 40 bugs from Defects4J's Lang and
Math projects that TBar is able to patch~\cite{liu2019:tbar} \emph{when given an idealized 
suspicious locations list that consists of only the location of the human-written patch}.
To this set, we add a few additional bugs that TBar was able to patch in our environment.
From this set, our modified process fails to patch the eight bugs at the bottom
of the table.
For two, the fault locater exhausts its five-hour time allotment.
For the other six, the slicer is essentially too aggressive and removes code
needed by TBar to recompile the file containing the failing test.
This code is unrelated to the failing test.
Because it does not recompile the tests, the removal does not inhibit the slicer.
We solved this problem by hand for Lang-33, where the failing test had the 
\emph{Assertion Roulette} smell~\cite{vandeursen2001:refactoring}, by refactoring the single
failing test into four smell-free tests.
In future work, we hope to automate a fix for the other six similar bugs.

\subsection{Metrics}

\noindent 
The experiments employ six metrics, 
Source Lines of Code, \SLoC,
Test-Suite Size, \TSS, 
Bug Rank, \BR, 
Repair Time, \RT,
Number of Patch Candidates, \NPC~\cite{qi2013:using}, and
Number of Test Executions, \NTE.
\SLoC measures the non-comment, non-blank lines of code.
In contrast to simple line counts, \SLoC better indicates the amount of effort
involved when working with source code.
The removal of code through slicing obviates the need for some of the tests.
We capture this reduction in the size of the test suite, \TSS.
Ideally, having less unrelated code to consider should focus fault localization
and improve, \BR, the rank of the buggy statement in the list of suspicious
locations.
We report repair time, \RT, as the wall-clock time used by TBar to patch a bug.
All of the experiments were run on the same hardware making the wall-clock time
proportional to the CPU time thus we do not report both.
\NPC counts the number of patch attempts before a valid patch is found
and has become a common metric for evaluating APR tools~\cite{chen2017:contractbased, ghanbari2019:prapr, liu2020:efficiency}.
In our work, it is used to quantify the efficiency gain brought about by using the more
cohesive reduced program.

Finally, we augment the use of \NPC by introducing a new metric \NTE, 
the number of test executions performed before a patch is found.
We introduce this metric because a reduction in the size of the test suite is expected to improve \RT but not \NPC, 
so \NPC cannot be used to study this \RT improvement.
The use of \NTE parallels the use of \NPC to study the \RT improvement
that comes from the improved ranking of the faulty statement.
Our goal is that \NTE and \NPC together will accurately characterize the reduction in \RT. 

\begin{figure}
\begin{centering}

\scalebox{0.9}{
  \begin{tikzpicture}
    \node (Cooo) at (0,0) {\Cooo};
    \node (Coor) at (-3,-1) {\Coor};
    \node (Coso) at (-1,-1) {\Coso};
    \node (Coop) at (1,-1) {\Coop};
    \node (Csoo) at (3,-1) {\Csoo}; %
    \node (Cosr) at (-4,-2) {\Cosr};
    \node (Cosp) at (-2,-2) {\Cosp};
    \node (Csor) at (0,-2) {\Csor}; %
    \node (Csso) at (2,-2) {\Csso}; %
    \node (Csop) at (4,-2) {\Csop}; %
    \node (Cssr) at (-1,-3) {\Cssr};
    \node (Cssp) at (1,-3) {\Cssp};

    \draw (Cooo) -- (Coor);
    \draw (Cooo) -- (Coso);
    \draw (Cooo) -- (Coop);
    \draw (Cooo) -- (Csoo);

    \draw (Coor) -- (Cosr);
    \draw[red] (Coor) -- (Csor);

    \draw (Coso) -- (Cosr);
    \draw (Coso) -- (Cosp);
    \draw[red] (Coso) -- (Csso);

    \draw (Coop) -- (Cosp);
    \draw[red] (Coop) -- (Csop);

    \draw[red] (Csoo) -- (Csor);
    \draw[red] (Csoo) -- (Csso);
    \draw[red] (Csoo) -- (Csop);

    \draw (Cssr) -- (Cosr);
    \draw (Cssr) -- (Csor);
    \draw (Cssr) -- (Csso);

    \draw (Cssp) -- (Cosp);
    \draw (Cssp) -- (Csop);
    \draw (Cssp) -- (Csso);
  \end{tikzpicture}
}
\end{centering}
\caption{The lattice of all possible configurations of program (\PP or \PS), test suite (\TT, or \TS), and suspicious locations list (\SL, \SLR, or \SLP).}
\label{fig:lattice}
\end{figure}

\subsection{Repair Process}

\noindent
Our repair process is applied to buggy program \PP and its test suite
\T, and consists of the four steps discussed below.
\begin{enumerate}

\item 
For the \emph{program reduction} step we use program slicing to produce \PR
from \PP by eliminating code irrelevant to the bug being repaired. 
    
\item 
The \emph{test-suite reduction} step reduces the test suite \T by removing
tests for code not found in \PR.
The result is a reduced test suite \TR that preserves \T's test coverage
relative to the bug being repaired.
    
\item 
The \emph{reduced suspicious list} is determined in one of two ways.
The first method regenerates the list by applying fault localization to \PR and \TR
producing \SLR, while the second method \emph{prunes} from the original list all locations 
that are not present in the slice, producing \SLP.

~~The goal of the first approach is to focus fault localization on the
relevant portions of the code (i.e., those found in the slice).
The advantage of the second approach is that the rank of the buggy statement is
guaranteed to be no worse.

\item 
The \emph{program repair} step applies TBar to various combinations of \PP,
\TT, \SL, \PR, \TR, \SLP, and \SLR, 
that conform the lattice shown in Figure~\ref{fig:lattice}, which is discussed below.
\end{enumerate}
To isolate the impact of the various reductions on TBar's performance, we
consider the lattice of possible configurations shown in Figure~\ref{fig:lattice}.
With two programs (\PP or \PS), two test suites (\TT, or \TS), and three suspicious locations lists (\SL, \SLR, or \SLP)
there are twelve possible configurations.
However, those shown in red are not viable because \TT and \SL include
program locations not found in the \PS.
In the lattice, each subscript indicates, in order, the program, the test
suite, and the ranked list of suspicious locations used.
Thus, \Cooo is the original configuration while \Cssp uses \PS, \TS, and \SLP.

\subsection{Tools}

\noindent
Our experiments use the following tools:
first, as previously mentioned, we use the APR tool TBar, available as part of the
replication package for the paper introducing TBar~\cite{liu2019:tbar}, 
as a representative state-of-the-art template-based APR tool.
The only modification required is to register the use of the sliced code
or reduced test suite with the version control system via \sfs{git add -u}
before applying TBar.

For Step 1, the reduction step, we use parallel ORBS version 5.0~\cite{islam2016:porbs}.
The test-suite reduction 
uses \emph{tree-sitter} to build the AST for each test file and then repeatedly
removes test methods that fail when applied to \PS or to \PP.
Step 3 uses the spectrum-based fault locater GZoltar V.1.7.3~\cite{campos2012:gzoltar}
with the Ochiai ranking strategy, 
which has empirically been shown to be one of the most effective metrics for fault localization~\cite{zou2021:empirical}.
Finally, the experiments were run in a Docker container on a cluster of AMD
EPYC 7763's.  Each run allotted TBar four processors and a time limit of 24 hours.

\subsection{Research Questions} 

\noindent
Using these tools we consider the following Research Questions.
We expect the overall improvement to come from two primary causes: improved
fault localization rank of the buggy statement and the reduced test suite.
RQ1 considers the impact of the reduction on the size of the buggy program.
RQ2 and RQ3 break out the two causes of improvement and attempt to quantify the
two reductions, while RQ4 and RQ5 consider their impact on TBar.
RQ6 considers the impact of our reduction techniques on TBar's repair time.

\begin{enumerate}[label={\bfseries\itshape RQ\arabic*:}, itemindent=1\parindent, leftmargin=1.8\parindent]

\item 
\textsl{How effective is slicing at reducing the size of the buggy program?}

Slicing has never been applied to preserving JUnit error messages and thus 
might prove ineffective at the task.
For this RQ we compare the \SLoC found in \PP and \PS.

\item 
\textsl{How effective is the reduced program when used for test-suite reduction?}

This RQ isolates the impact of using the reduced program to cull the test
suite, which we measure using \TSS.

\item
\textsl{How does program reduction impact fault localization?}  

This RQ isolates the impact of using the reduced program to improve the rank of
the buggy statement, which is measured using \BR in both \SLR and \SLP.

\item
\textsl{How does using a reduced test suite affect TBar?}

RQ4 builds on RQ2 to investigate the test-suite reduction's impact on TBar's
testing effort, as measured using \NTE.

\item
\textsl{What is the impact of using the updated list of suspicious statements 
on TBar?} 

RQ5 builds on RQ3 to investigate the improvement better bug rank (lower \BR) 
has on TBar's performance as measured using \NPC.

\item
\textsl{What is the overall improvement obtained using TBar with the reduced program?}

Finally, we use \RT to study the end-to-end performance improvement that the reduced
program has on TBar at various points in the lattice shown in Figure~\ref{fig:lattice}.

\end{enumerate}

\section{Results and Discussion}

\noindent
This section considers each of our research questions in turn and concludes with a
discussion of the implications of the patterns seen in the data.

\subsection{RQ1}
\begin{figure}[b]
\centering
\includegraphics[width=0.5\textwidth]{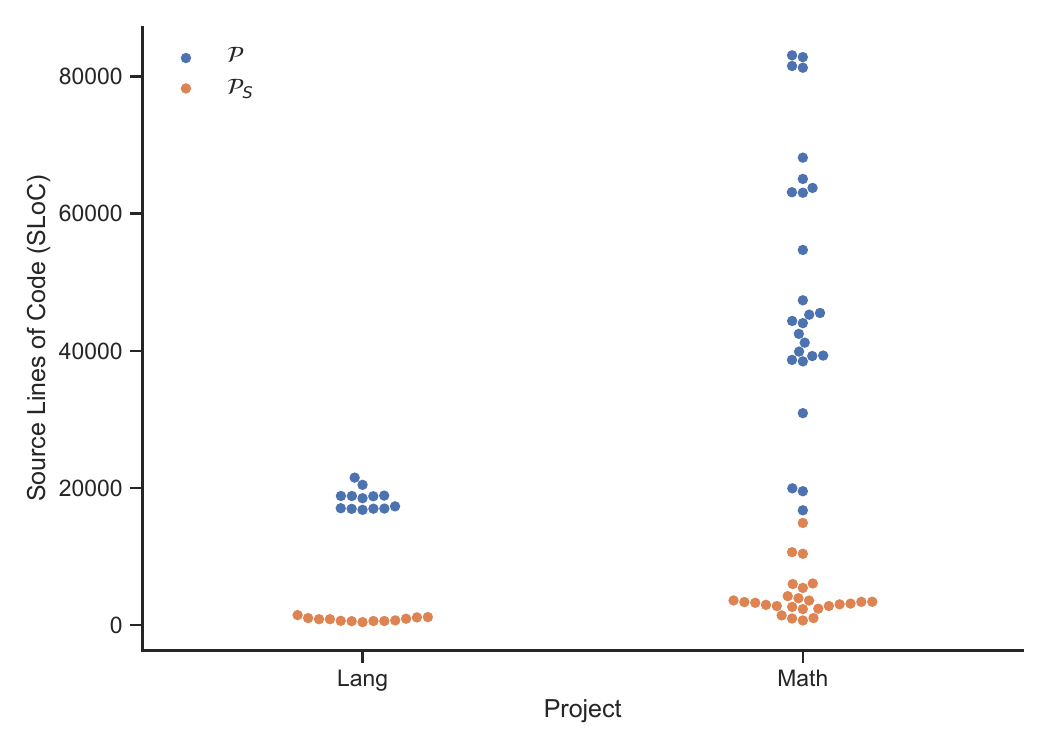}
\vspace*{-4.5ex}
\caption{Reduction's impact on each program's \SLoC.}
\label{fig:RQ1}
\end{figure}

\noindent
RQ1 considers the size reduction achieved using program slicing.
Figure~\ref{fig:RQ1} shows the size of the production code, which excludes test
code and alike, compared to the size of the reduced production code.
As noted previously, slicing has never been applied to preserving JUnit error
messages.
While there are interesting issues such as Math-5's reciprocal test issue
discussed with RQ3, overall the approach is quite effective.
Compared to the average dynamic slice size of approximately
60\% of the code~\cite{binkley2019:comparison},
for the programs shown in Table~\ref{tab:subjects} the slices range from 2.2\%
to 18.3\% of the original code with an average size of only 6.6\%.
Some of this is likely related to the nature of the particular subjects, but
using JUnit tests as a slicing criterion may prove worthy of future study.
It would be interesting to do a more apples-to-apples comparison using 
subjects studied in prior dynamic slicing research.

\begin{mdframed}[style=mystyle]
In answer to RQ1, slicing is quite effective at
reducing the size of the program necessary to reproduce the bug.
Furthermore, this reduction is fairly uniform.
\end{mdframed}

\subsection{RQ2}

\begin{figure}[b]
\centering
\includegraphics[width=0.5\textwidth]{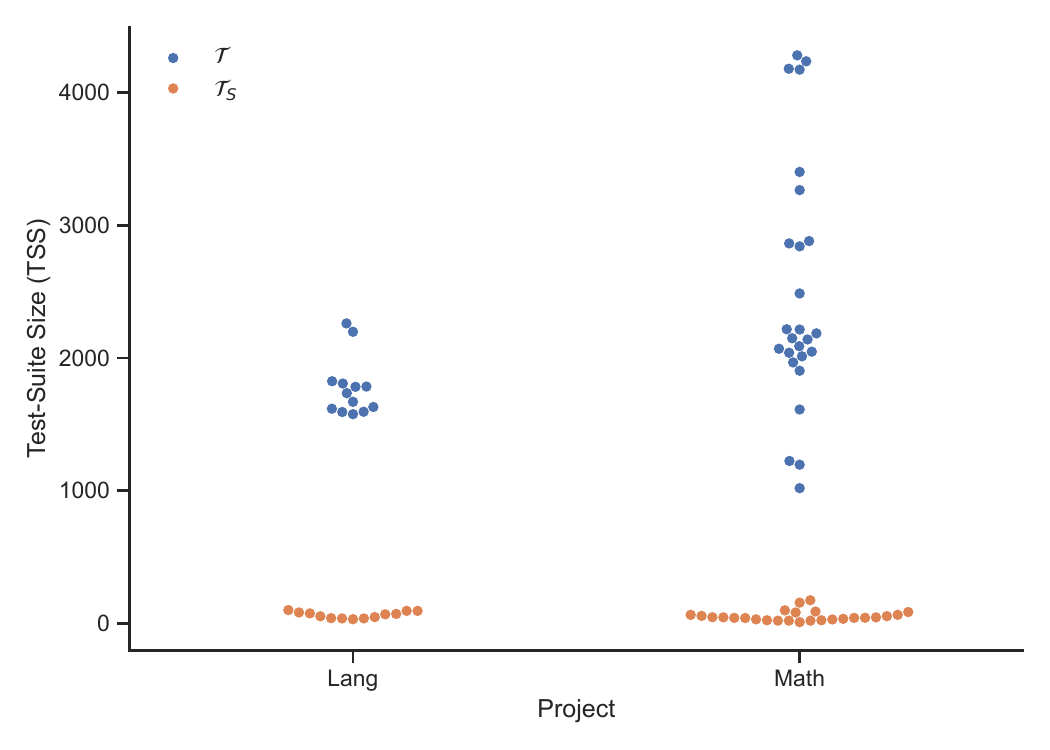}
\vspace*{-4.5ex}
\caption{Reduction's impact on each program's \TSS.}
\label{fig:RQ2}
\end{figure}

\noindent
RQ2 considers the test-suite size reduction obtained by removing tests that
apply to parts of the code not in the slice.
Here again, as seen in Figure~\ref{fig:RQ2} as well as
Table~\ref{tab:subjects}, using the reduced program leads to a dramatic reduction 
in \TSS.
The original test suite includes an average 2249 tests per project, which is
reduced to an average of only 56!
The greatest reduction, for \sfs{Math-89}, leaves only eight passing and one
failing test.
The magnitude of the reduction is, de-facto related to the size of the
original test suite.
It would be interesting to study the reduction after applying traditional
test-suite minimization~\cite{cruciani2019:scalable}.

One interesting question the dramatic reduction raises is, is the reduction too
extreme.  
That is does it not leave sufficient tests for the downstream tools to work
effectively?
In theory TBar needs only a single failing test case.  
Of course, in the absence of good regression tests, overfitting is a
potential problem.
At the same time many of the tests in the original test suite have zero impact
on the repair process and thus their repeated execution is a significant waste
of time.

\begin{mdframed}[style=mystyle]
In answer to RQ2, slicing is quite effective
reducing test-suite size.
\end{mdframed}

\subsection{RQ3}

\noindent
Starting with RQ3 we remove from the dataset those projects that TBar fails to
patch.
The initial set of projects we started with includes all projects that TBar can
patch \emph{given an ideal suspicious statements list of length one}~\cite{liu2019:tbar}.
In less ideal situations (e.g., using GZoltar's list of suspicious statements
in lieu of the ideal list) there are cases where TBar is unable to find a patch.
These bugs are marked with a ``*'' in Table~\ref{tab:subjects}.

RQ3 investigates \BR, the rank in the list of suspicious statements of
the statement to which TBar applies a successful patch.
It is important to note that we are not reporting the location of a
predetermined static statement but rather the statement dynamically chosen by
TBar.

The predetermined static statement approach is clearly more straightforward:
ahead of time, one selects a statement to be the correct patch location and then
determines its rank on each suspicious statements list.  
In cases where there is only one correct patch location, this approach has
some appeal.
It also simplifies correctness arguments when, for example, the predetermined
location is taken from a subsequently patched version of the code.
However, this approach denies the possibility of alternate fixes at alternate
fix locations and thus is rather limiting.

The dynamic approach is more flexible as it gives TBar credit for finding
patches at alternative locations.
Of course, alternative locations complicate the correctness argument.
Because of this, we consider two sets of bugs.
The first, \emph{all the data} includes those bugs for which TBar finds a patch
in the base \Cooo configuration.
These are listed in Table~\ref{tab:subjects} without a ``*''.
We use these bugs to establish overall trends.
The second set considers only those projects where the patch location is the
same as that found using \Cooo.
As with the straightforward approach, this enables us to make direct
correctness arguments.
One interesting question here is whether this subset shows the same general trends.

For both \emph{all the data} and the subset using the same patch location, our
analysis considers two different reduced suspicious statements lists, \SLP and
\SLS.
By removing statements from \SL that are not in the slice, \SLP has the
advantage that it is never larger than \SL.
Assuming that location patch-ability does not change, using \SLP should enable
TBar to choose a statement of lower or equal rank.  
As discussed below, this assumption is violated in one case.

The second list, \SLS, hopes that GZoltar does better when applied to the more
focused \PS and \TS.  
It mostly does.
In the data, there are only two cases, Math-5 and Math-75, where the list of
suspicious statements produced by GZoltar is larger for \SLS than for \SL.
We leave it to future work to better understand if these longer lists provide
evidence of GZoltar finding a greater number of viable locations or if the
longer lists simply include more false positives.

\begin{figure}
\centering
\includegraphics[width=\linewidth]{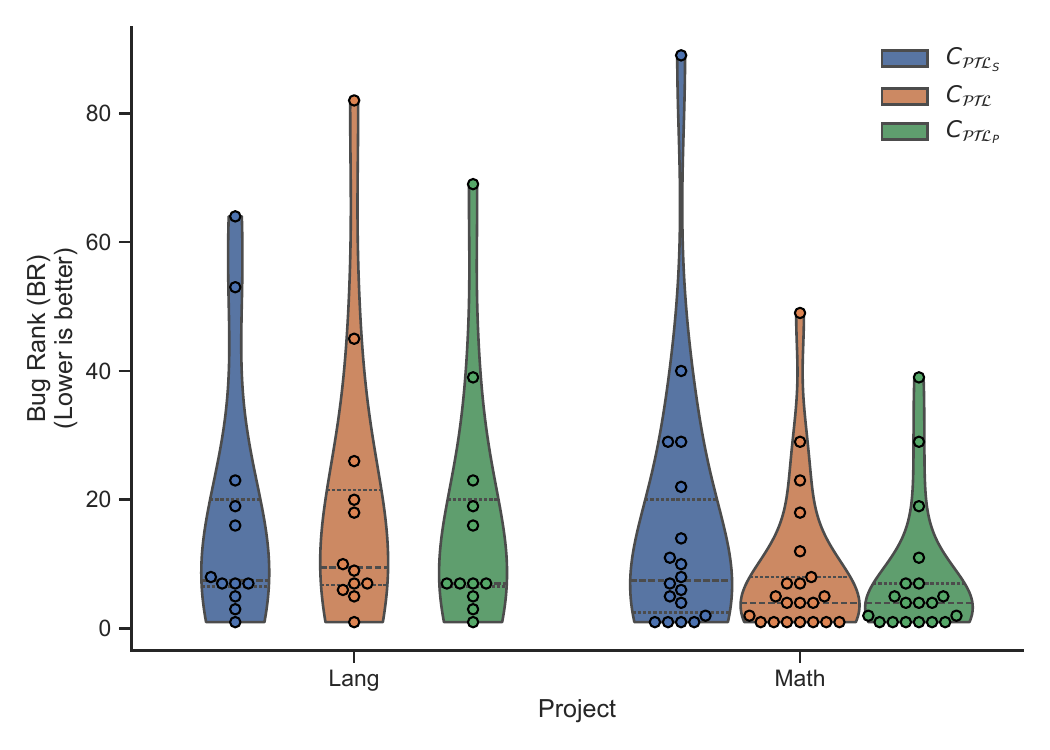}
\includegraphics[width=\linewidth]{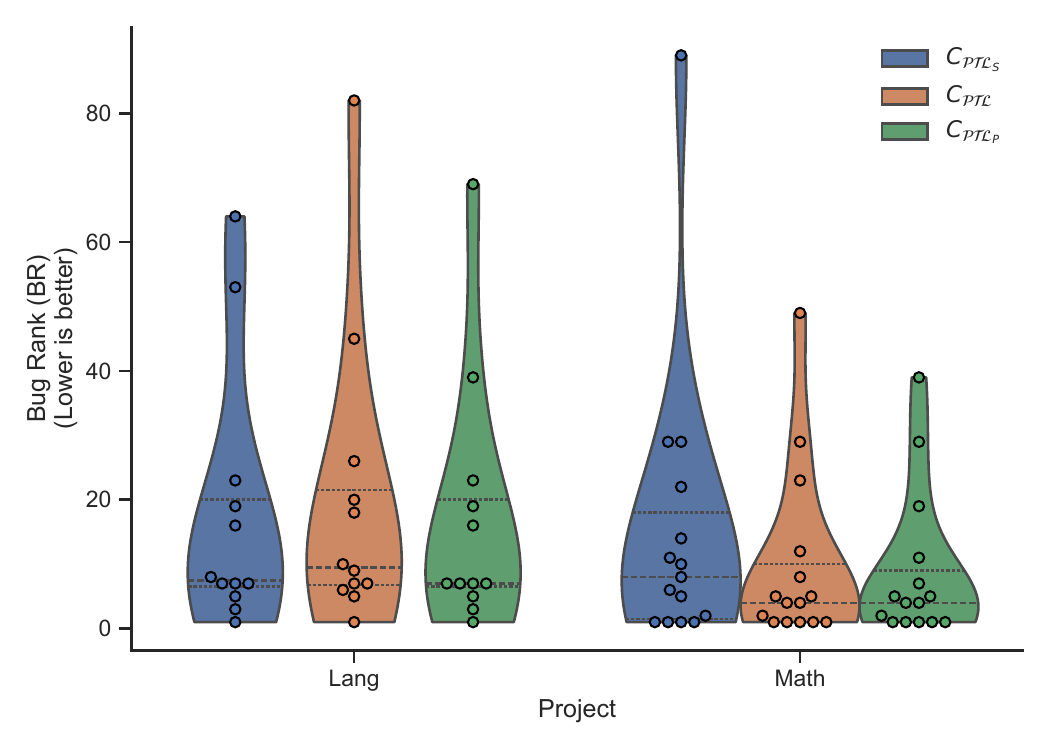}
\vspace*{-4.5ex}
\caption{Reduction's impact on each program's \SL.}
\label{fig:RQ3}
\end{figure}

Turning to the \BR data, we first consider all of the data.
Showing the baseline \SL in the middle, the upper chart of Figure~\ref{fig:RQ3}
compares \BR separately for Lang and Math.
TBar is unable to patch one bug, Math-6, under \Coop
and three, Math-6, Math-35, and Math-64 under \Coor.
Thus from an applicability perspective \Coop is preferred.

The expected pattern, where the reduced list produces lower \BR{}s, is mostly
evident in the figure.
For example, the max and median \BR{}s using \SLP (shown to the right of \SL, in
green) are lower than in \SL (the median is shown as the central dashed line).
(Numerically, these are 69.0 and 7.0 verses 82.0 and 9.5.)~
For \SLS, Lang shows this same pattern, but Math does not, largely because of
one bug discussed below.

Considering the individual bugs, both Lang and Math include a handful of
bugs for which \BR in \SLS is larger than \BR in \SL.
Most of the time the difference is small, but not always.
For example, Math-82's \BR is 89 in \SLS while only 49 in \SL.  
This one project accounts for the long thin spire of the \SLS Math violin plot
and dominates the reason that Math does not follow the expected pattern.

For \SLP, pruning \SL can never make the list longer.  
It is therefore easy to likewise conclude that the \BR of the patched
statement can never be greater using \SLP.
However, checking the data, we find one exception, Math-5 where the \BR in \SL
is one while the \BR in \SLP is two.
For this to happen, the rank-one statement from \SL cannot be in the slice and
thus gets removed during pruning.
That another viable patch location was found is interesting.
Fortuitously, it was not at rank one, otherwise this anomaly might have gone
unnoticed.
The details are explored in the discussion at the end of this section.

The lower chart in Figure~\ref{fig:RQ3} shows only those bugs where TBar
patches the same location using the three lists. 
For Lang all the bugs use the same location, thus the charts are identical.
For Math fifteen of the bugs use the same location, while six are patched at different
locations.
The dominance of same-location patches means that most patches inherit patch
correctness from the correctness of the \Cooo patch.
However, a different patch location does not mean the patch is incorrect,
it simply means that it is harder to argue its correctness.  
Finally, as expected for this subset, \SLP always results in a patch location with an
equal or lower \BR.

\begin{mdframed}[style=mystyle]
To answer RQ3, while for the most part the rank
of the statement that TBar patches is lower in \Coop and \Coor than in \Cooo,
this is not ubiquitous, nor is the reduction as dramatic as that seen
for \SLoC and \TSS.
It is encouraging that most patches are at the same location, which implies
that the reduction in rank does not negatively impact the patch.
Finally, overall \Coop performs better than \Coor.
\end{mdframed}

\subsection{RQ4}    %

\begin{figure}[b]
\centering
\includegraphics[width=\linewidth]{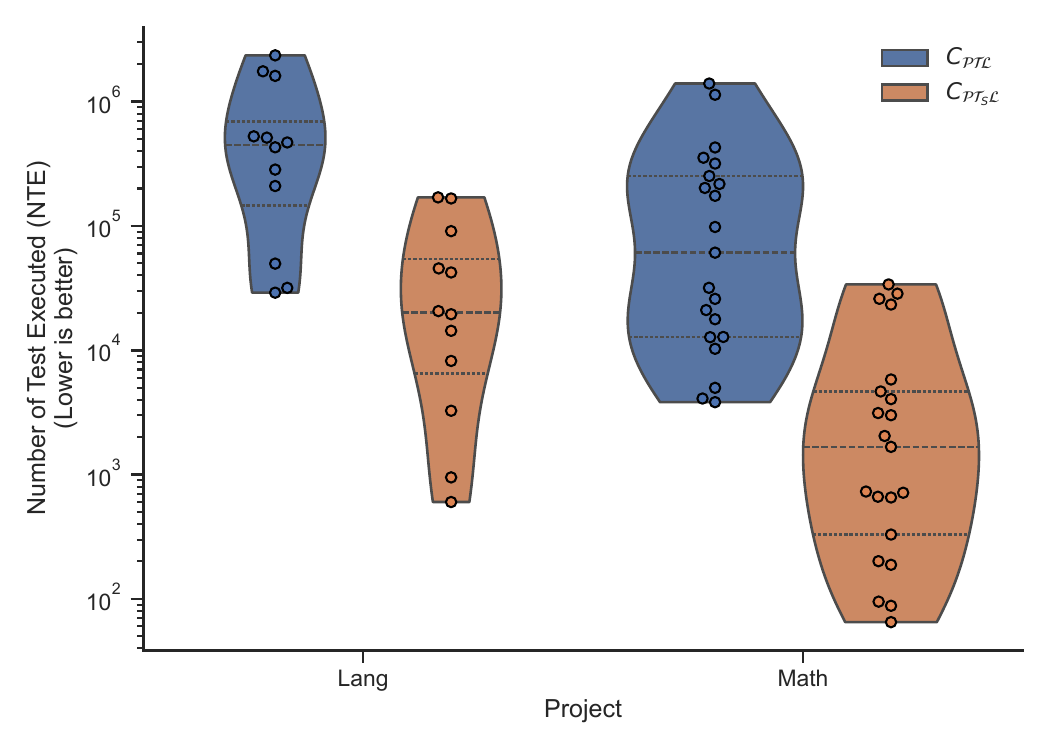}
\vspace{-5ex}
\caption{Reduction in test-suite size's impact on each project's \NTE.}
\label{fig:RQ4}
\end{figure}

\noindent
RQ4 investigates the effort TBar expends testing potential patches, measured
using the number of test executions, \NTE.
Similar to our analysis of RQ3 we first consider all the data and then the
subset where \Cooo and \Coso choose the same patch location.
Figure~\ref{fig:RQ4} compares the testing effort 
using all of the data.
Note that the $y$-axis uses a logarithmic scale.

The reduction is dramatic.
For Lang, \NTE drops from an average of 687,946 to an average of only 48,492, a
93\% reduction.
For Math, \NTE drops from 227,604 to 6,649, a 97\% reduction.
Clearly, the reduction in \TSS seen in RQ2 leads to a dramatic reduction in 
\NTE.  

Even better, unlike in RQ3, where using the reduced suspicious statements
lists, caused TBar to fail to patch three of the bugs, when using the reduced
test suite, TBar is able to patch all of the bugs.
On top of this, it does so at the same location; thus Figure~\ref{fig:RQ4}
is also the figure when using the same-location subset.

\begin{mdframed}[style=mystyle]
To answer RQ4, slicing is successful at reducing TBar's testing
effort, \NTE, without reducing its applicability.
\end{mdframed}
\vspace{-2ex}

\subsection{RQ5}

\begin{figure}[b]
\centering
\includegraphics[width=\linewidth]{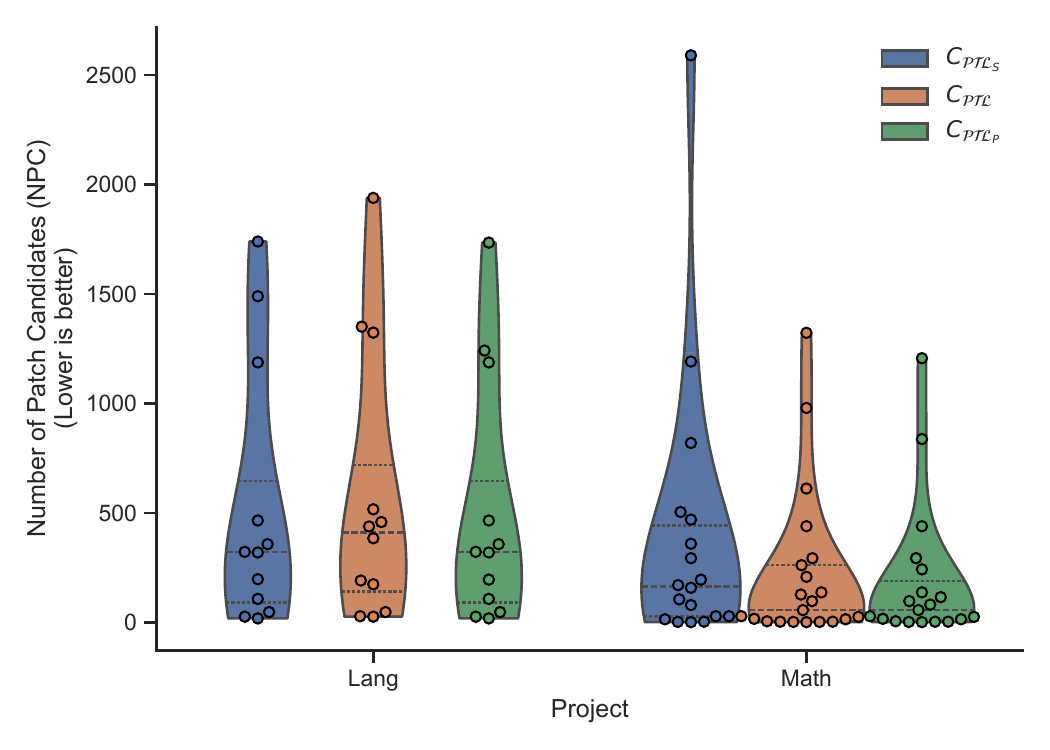}
\vspace*{-1ex}
\includegraphics[width=\linewidth]{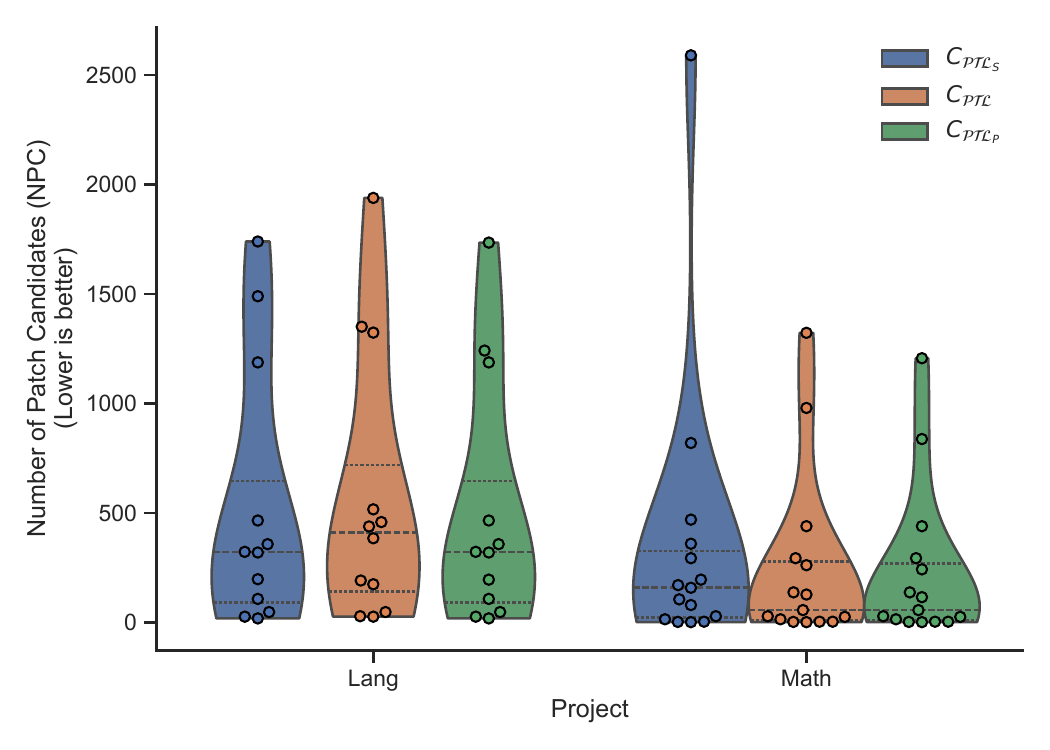}
\vspace{-5ex}
\caption{Reduction's impact on each program repair's \NPC.}
\label{fig:RQ5}
\end{figure}

\noindent
RQ5 mirrors RQ4 replacing \NTE with \NPC, the number of patch candidates. 
The impact on NPC of using the updated list of suspicious statements is shown
in Figure~\ref{fig:RQ5} where it is visually obvious that the improvement is
less dramatic than for \NTE.
Comparing the center violin  plot with its right and left neighbors, using \SLP
always brings a benefit.
The same is not true for \SLS where the inferior ranks for \SLS seen in RQ3
are again evident for Math.

For the pruned list \SLP, average \NPC drops from 573 to 502 for Lang, a 12\%
reduction, and from 220 to 189 for Math, a 14\% reduction.
Using \SLS the results are mixed.
For Lang \NPC drops from 573 to 523, a 9\% reduction, but for Math it rises
from 220 to 389, a 77\% \emph{increase}.
The inferior and varied performance attained using \SLS stands in contrast to
the expectation that performing fault localization on the reduced program would
focus the fault locator and thus improve its performance.
The data suggests that this is not the case.
It would be interesting to investigate the interplay between the amount of code
reduction and the performance of fault localization.

Turning to the subset where the same patch location is used, because all
of the Lang patches are at the same location, the results for the same-location
subset are identical.
For Math the performance is mixed.
Using \SLP \NPC drops from 246 to 227, which is only an 8\% reduction compared
to the 14\% for all the data.
On the other hand, while still negative, for \SLR the increase from 246 to 352
is only a 43\% increase, down from the 77\% when using all the data.

\begin{mdframed}[style=mystyle]
In answer to RQ5, as expected the data mirror the
\SL-data from RQ3.
Compared with the \NTE results for RQ4, slicing is notably less effective at
reducing \NPC.
\end{mdframed}

\subsection{RQ6}

\begin{table}
\centering
\small
\begin{tabular}{l| r r r c}
\multicolumn{1}{c|}{Configuration} &
\multicolumn{1}{c}{All} &
\multicolumn{1}{c}{Lang} &
\multicolumn{1}{c}{Math}  &
\multicolumn{1}{c}{Failures} \\

\midrule

\multicolumn{5}{l}{Three changes} \\
~~~~\Cssp  &  82\%     & 91\%    &   74\%    & 11  \\
~~~~\Cssr  &  78\%     & 88\%    &   67\%    & 12  \\
\midrule
\multicolumn{5}{l}{Two changes} \\
~~~~\Cosp  &  88\%     & 89\%    &   87\%    &  ~2  \\
~~~~\Cosr  &  83\%     & 89\%    &   76\%    &  ~3  \\
\midrule

\multicolumn{5}{l}{One change} \\
~~~~\Coso  &  86\%     & 86\%    &   85\%    &  ~0  \\
~~~~\Coop  &  14\%     & 15\%    &   12\%    &  ~2  \\
~~~~\Coor  & -75\%     &  7\%    & -174\%    &  ~3  \\

\end{tabular}
\vspace{2ex}
\caption{Repair time reduction relative to \Cooo. 
The final column reports the number of bugs where TBar succeeds using \Cooo
but not the listed configuration.}
\label{tab:RT}
\end{table}

\noindent
RQ6 investigates the impact of combining the reductions of \NTE and \NPC on
TBar's repair time, \RT.
The greatest reduction is expected with either \Cssp or \Cssr, which make use
of the reduced program, the reduced test suite, and one of the reduced
suspicious statements lists.  
Overall, these configuration yield almost an order of magnitude reduction, 
as can be seen in the top two rows of Table~\ref{tab:RT}, which breaks the
repair time reduction down by project and also reports the number of bugs that
TBar fails to patch despite \Cooo succeeding. 
Breaking this down by project, for Lang using \SLP the mean repair time drops
from 10946s to 990s, a 91\% reduction, while for \SLS it drops to 1290, an 88\%
reduction.
The reduction for Math is less dramatic.
For Lang the \SLP reduction, from 8324s to 2251s, is 74\%, while for \SLS
the reduction to 2701s, is 67\%, which is still notable.
The drawback in these impressive reductions is the number of TBar failures.
Using \PS, TBar is unable to patch roughly a third of the bugs that it can
patch using \PP.
That TBar struggles to patch the smaller sliced program comes as a bit of a
surprise and warrants future investigation.

To better understand the reduction and the failures, we consider the
intermediate points of the configuration lattice shown in
Figure~\ref{fig:lattice}.
Starting with the second level of Figure~\ref{fig:lattice}, as shown in the last
three lines of Table~\ref{tab:RT}, test-suite reduction clearly brings the
larger benefit.
In contrast the individual reduced suspicious statements lists bring mixed
benefit.
In the worst case, using \SLR actually hurts the performance of the Math bugs.
The negative performance is dominated by Math-5 whose repair time goes from 492s to
31671s. %
In contrast, the \SLR lists improve Lang's repair time by a modest 6\%.

With Lang, \SLP fairs better, seeing a respectable 14\% reduction.
However, the ``reduction'' for Math is again negative.
It is notable that \SLP leads to an average reduction of 12.6\%
when Math-5 is excluded.
We return to Math-5 in the discussion section, 
since this anomaly %
warrants deeper investigation.

The center of Table~\ref{tab:RT} considers the level in the lattice where
two of the three possible reductions are applied.
Here the test-case reduction's impact again clearly dominates.
It is interesting that compared to \Coso, \Cosp provides a strict
improvement for Lang and Math.
This gives us optimism that with additional effort we can devise suspicious
statements list reduction techniques that are effective across all projects.

\begin{figure}
\centering
\includegraphics[width=\linewidth]{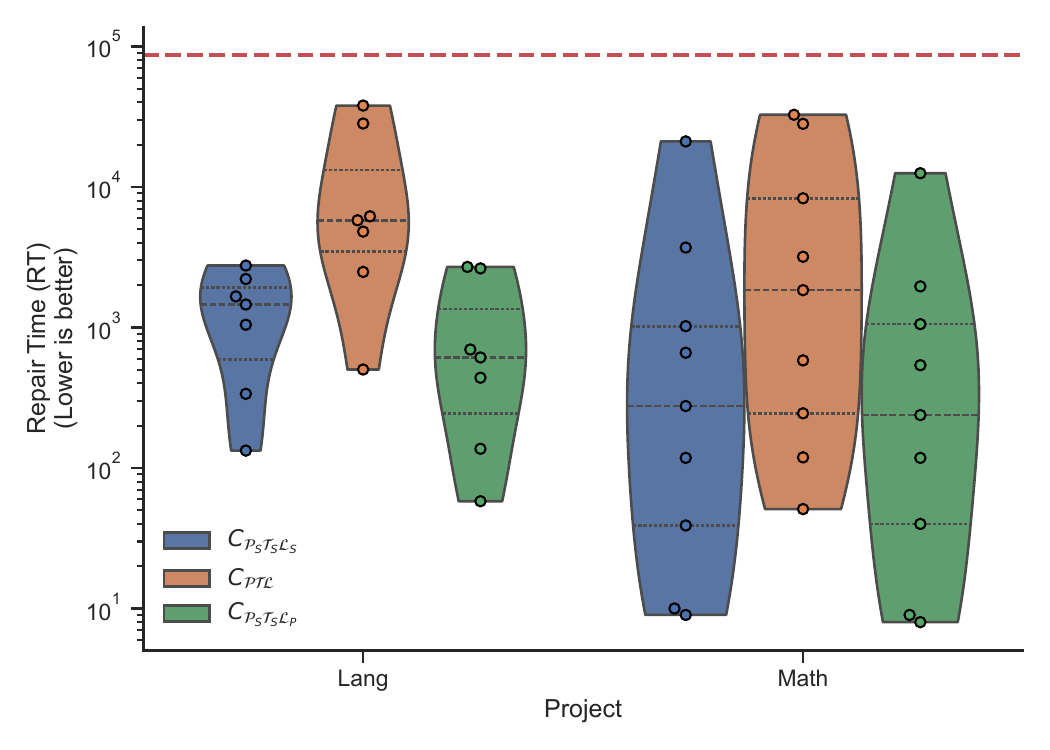}
\vspace{-4.5ex}
\caption{Reduction's impact on overall repair time, \RT, for \Cssr and \Cssp restricted to the bugs where TBar chooses the same patch location.}
\label{fig:RQ6l3}
\end{figure}

\begin{figure}
\centering
\includegraphics[width=\linewidth]{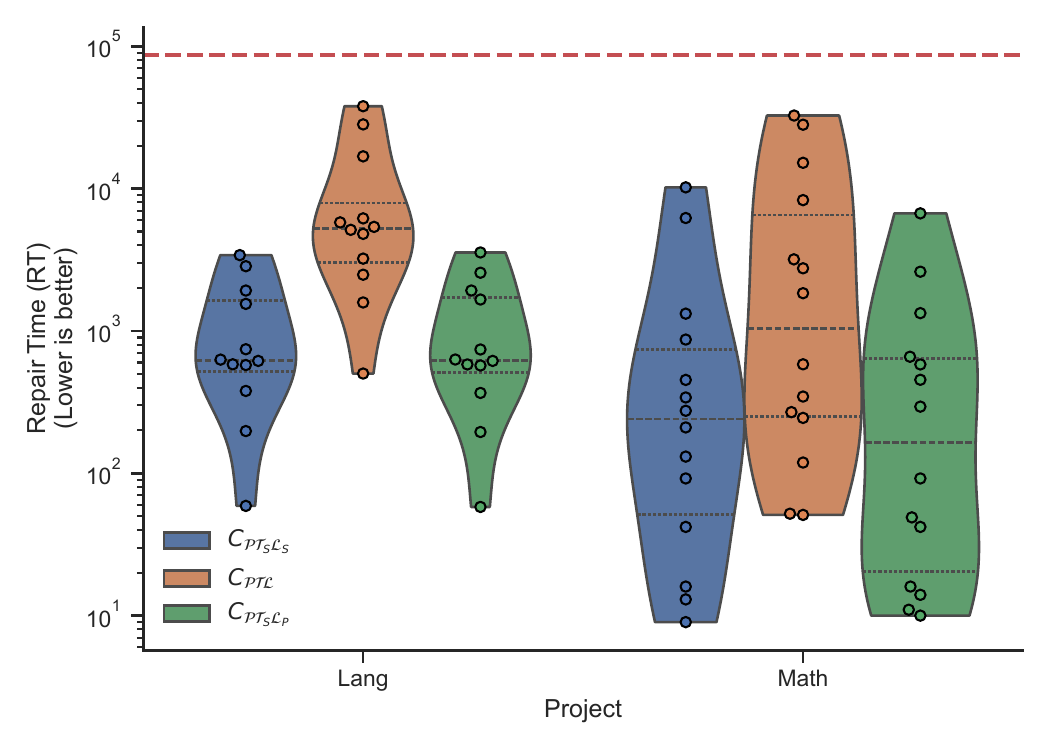}
\vspace{-4.5ex}
\caption{Reduction's impact on overall repair time, \RT, for \Cosr and \Cosp restricted to the bugs where TBar chooses the same patch location.}
\label{fig:RQ6l2}
\end{figure}

Comparing the two middle and the top two rows, using \PS benefits
only one case, with Lang using \Cssp, which is a bit discouraging.
Much of the difference is caused by bugs where TBar uses different patch
locations.
To dig deeper, Figures~\ref{fig:RQ6l3} and~\ref{fig:RQ6l2} 
compare the performance on the subset where TBar uses the same patch location.
Figure~\ref{fig:RQ6l3} compares \Cssr and \Cssp,
while Figure~\ref{fig:RQ6l2} compares \Cosr and \Cosp.
The two figures show very much the same reduction pattern, except that
Figure~\ref{fig:RQ6l2} summarizes 50\% more data, which underscores the
challenge TBar faces trying to patch \PS.

Finally, it is interesting that despite \PS and \TS showing similar large
reductions, the \TS reduction has a dramatic impact on \RT while the \PS
reduction does not.
The obvious challenge here is to find a way for \PS to have a similar
impact.

\begin{mdframed}[style=mystyle]
In answer to RQ6 the results are somewhat mixed.
Clearly the \TS reduction brings \RT benefit.
However, the \PS reduction does not.
The overall impact is still impressive, but would be even more impressive if
future APR tools could better leverage the \PS reduction.
\end{mdframed}

\subsection{Discussion}
\label{sec:discussion}

\noindent
This section reflects on some of the questions raised in the experiments.
To begin with, the bottom of Table~\ref{tab:subjects} includes eight bugs
that failed before getting to the TBar stage of our tool pipeline.
For two of these GZoltar times out.
So, despite the fact the TBar can patch these two given the location of the
bug, finding that location proves a challenge.
One interesting question here is why GZoltar was not able to leverage the more
focused \PS and \TS.  

For the other six bugs, the slicer is essentially too aggressive and
removes code needed by Defects4J to compile the test suite. 
(TBar works through the Defect4J interface.)
For efficiency the slicer does not recompile the \sfs{Java} test files.
This means that if there is test code in the same file as the failing test,
the slicer may remove production code needed to compile that test even though
it is never executed.
An ad-hoc preliminary experiment that modified ORBS to verify that the test
file compiled solved this problem in one case.
We plan to work on better engineering solutions that more generally address
this issue.

Second, compared with the \NTE results for RQ4, slicing is notably less
effective at reducing \NPC and \RT. 
In theory the search for locations in \PS should be more efficient, so here
again slice-aware fault location might be interesting to explore.
Since most of the slices include the patch location, it seems
reasonable to devise fault locators that can leverage being applied to the
slice.
Another consideration is that Lang and Math are both utility collections.
It is possible that for a more monolithic program, \PS would have greater
impact.
In terms of \RT, one experiment worth running is to consider a project that
involves computationally complex aspects.
If such code is not in the slice then \PS should lead to \RT improvements 
akin to those \TS provides.

While the test-suite reduction is impressive it is possible that insufficient
tests remain.
Math-6 had by far the most failing test cases (and the second highest number of
passing tests in \TS), so it is worthwhile to consider if it was particularly
easy or challenging to patch.
For starters, the rank of the patch location in \SL is one.  
An optimistic reading of this is that all the failing tests helped locate the
fault.
Unfortunately plenty of other patches involve the Rank 1 location.
Math-6 is one of two bugs that \Coor is unable to patch, which suggests
that insufficient tests is not limiting GZoltar.
Comparing  our other metrics, none stand out.
Perhaps the silver lining here is that patch success does \emph{not} depend on
having a large number of failing or passing tests.

Finally, we consider two cases studies, Lang-33, which exhibits the Assertion
Roulette test smell~\cite{vandeursen2001:refactoring}, and Math-5, which was successfully
patched at an alternate location.
To begin with, Figure~\ref{fig:lang33} shows a code excerpt in which the test
method \lsin{testToClass_object} includes multiple asserts and thus exhibits
the Assertion Roulette test smell~\cite{vandeursen2001:refactoring}.
Two of the test's four asserts are shown (on Lines 2 and 4).
Following the test method, Lines 8-19 show the method under test.
This code generates a \sfs{null dereference exception} on Line 16 when
\lsin{i=1} because \lsin{array[i]} is \lsin{null} for the test on Line 5.

The slicer unwantedly removes Lines 9-13 from the method \lsin{toCLass} without
affecting the program behavior because it still generates a \sfs{null
dereference exception}. 
However, the exception is now coming from the use of \lsin{array.length} on
Line 14 and is now generated by the assert on Line 2.
Lines 9-13 get removed because their absence does not change the
program behavior.
Refactoring the test method into four test methods, and thus removing the
Assertion Roulette test smell, fixes the issue because the initial deletion of
Lines 9-13 now causes the first test to fail \emph{in addition} to the fourth.
This example highlights one of the challenges caused by using bug error
messages as a slicing criterion.
The results for Lang-33 reported in this paper make use of the refactored test.

\begin{figure}
\begin{lstlisting}[language=Java]
public void testToClass_object() {
  assertNull(ClassUtils.toClass(null));
  ...
  assertTrue(Arrays.equals(... 
    ClassUtils.toClass(... { "Test", null, ....
}

public static Class<?>[] toClass(Object[] array) {
  if (array == null) {
    return null;
  } else if (array.length == 0) {
    return ArrayUtils.EMPTY\_CLASS\_ARRAY;
  }
  Class<?>[] classes = new Class[array.length];
  for (int i = 0; i < array.length; i++) {
    classes[i] = array[i].getClass();
  }
  return classes;
}
\end{lstlisting}
\vspace*{-2ex}
\caption{Example from Lang-33.}
\label{fig:lang33}
\vspace*{-2ex}
\end{figure}

\begin{figure}
\begin{lrbox}{\mylisting}
\begin{lstlisting}[language=Java]
// assertion method used: assertEquals(expected, actual)

// Original error output
// Test failed!
// testReciprocalZero(... .math3.complex.ComplexTest): 
//   expected:<(NaN, NaN)> but was:<(Infinity, Infinity)>

// Correct error output
// testReciprocalZero(... .math3.complex.ComplexTest): 
//   expected:<(Infinity, Infinity)> but was:<(Nan, Nan)>

// the test
  public void testReciprocalZero() {
    assertEquals(Complex.ZERO.reciprocal(), Complex.INF);
  }

// Original patch replaces 
  if (real == 0.0 && imaginary == 0.0) {
    return NaN;
  }
// with
  if (real == 0.0 && imaginary == 0.0) {
    return INF;
  }

// Sliced code under test
  } else {
    double q = imaginary / real;
    double scale = 1. / (imaginary * q + real);
    return createComplex(scale, -scale * q);
  }

  // Math definition if INF (infinity)
  public static final Complex INF 
    = new Complex(Double.POSITIVE_INFINITY, 
                  Double.POSITIVE_INFINITY);

  public boolean equals(Object other) 
  ...
replacing
   return (real==c.real) && (imaginary==c.imaginary);
with
   return ((real==c.real) && (imaginary==c.imaginary)) 
           || (isNaN || Double.isInfinite(imaginary));
\end{lstlisting}
\end{lrbox}
\resizebox{\columnwidth}{!}{\usebox{\mylisting}}
\vspace*{-1ex}
\caption{Example from Math-5.}
\label{fig:math5}
\vspace*{-2ex}
\end{figure}

The second case study looks at the Math-5 anomaly.
Although pruning \SL to produce \SLP should not result in the use of a lower-ranked location as it cannot make the list longer, 
our investigation of RQ3 found that this happened for Math-5, as the Rank 1 location was pruned away from \SL.
The cause is complex but illustrative.
Figure~\ref{fig:math5} shows excepts from the relevant code.
To begin the assertion method used (Line 1 of Figure~\ref{fig:math5}) takes as
its first argument the expected value.
The call, shown on Line 14,
inverts \lsin{expected} and \lsin{actual}.
It should use \lsin{Complex.INF} as the first argument.
While this does not impact the slicer nor TBar, which both only look for
differences in the output, it does impact human understanding.
Specifically, the test output shown on Lines 5 and 6 
is backwards.  
The correct output is shown on Lines 9 and 10.
Again the slicer and TBar only monitor change, so the slice and the patch are
correct but the order is relevant to understanding the interplay between \SLS
and \SL. 
Using \Cooo, the original TBar patch replaces Line 19 with Line 23.
This patch correctly returns the reciprocal of \lsin{0} as the library's
constant \lsin{INF}. 

Using \Coop, the slice omits the original patch location (Lines 18-20), which
means the location is pruned from \SL when producing \SLP and thus is not
available to TBar.
The slicer can remove Lines 18-20 because subsequent code (shown on Lines 27-30)
results in a division by zero when \lsin{real} is \lsin{0}, which the \sfs{Java}
runtime maps to the value infinity.
The value returned by \lsin{createComplex} on Line 30 is
\lsin{Complex(Infinity, Infinity).}
It is important to note that this value is \emph{not} the same value as
the constant returned on Line 23, whose declaration is shown on Line 34.
However, the two \emph{print} the same; thus, the deletion leaves the error
message unchanged.
Detecting no change in the output, ORBS concludes that the relevant program
behavior has not changed.
While it may not appear serious, this is an informative limitation of using
value-based slicers such as ORBS.
Interestingly, if the program had printed \lsin{INF} differently from Java's
default printing of infinity, the deletion would not have occurred.

Unfortunately, the patch TBar produced under \Coop is inferior, or
this would have made a great example of the value of considering
alternate patch locations.
The patch, shown on Line 43, modifies the \lsin{equals} method so that
all complex numbers equal NaN, which is clearly undesired.
Still, the fact that another viable patch location was found is interesting
and suggests that APR's use of a ``perfect patch'' may be undesirable.

 ORBS is expensive.
 It is possible that other, unsound, lighter-weight reduction techniques might
 be sufficient, but to understand the tradeoffs, we first need to understand how
 effective ORBS is when applied to the buggy code.

Finally, our experiments make use of one-minimal ORBS slices. 
Because these are minimal and, by construction, capture the execution of the
slicing criterion, their use allows us to focus on slicing's impact.
This is akin to the experiments with TBar that use an optimal list of
suspicious statements~\cite{liu2019:tbar}.
In practice, ORBS is too slow to be used in production (while some engineering
work would reduce its runtime, the median slice time for Lang was just over
seven hours and that for Math was just under 34 hours).
Fortunately, there exist faster slicers.
For example, the slicer studied by Wang et al.~\cite{wang2020:type}
when slicing JFreeChart (an application similar is size to Lang) slices at
about 400 SLoC per second (including SDG construction time).
Thus, we expect it to slice the largest of the systems we consider in under five minutes.

\section{Related Work}

\noindent
We distinguish several areas of related work, 
ranging from techniques to improve APR and fault localization, 
to related applications of program reduction.

\head{Improving APR}
Early research in APR focused on developing new patch-generation algorithms
such as heuristic-based~\cite{legoues2012:genprog, tian2017:automatically},
constraint-based~\cite{afzal2021:sosrepair, mechtaev2018:semantic}, and
learning-based~\cite{chen2021:fast, gupta2017:deepfix}, all aimed at improving the
efficacy of APR by producing more correct patches and addressing the
well-known patch overfitting problem~\cite{qi2015:analysis, xiong2018:identifying}.

More recently, researchers have started focusing on techniques to improve the efficiency of APR.
Several techniques have been developed, 
including regression test selection~\cite{yoo2012:regression,mehne2018:accelerating}, 
patch filtering~\cite{liang2021:interactive,yang2021:accelerating},
and on-the-fly patch generation~\cite{hua2018:practical} and validation~\cite{chen2021:fast,benton2022:boosting}. 
The goal of these techniques is to reduce the cost of patch compilation and test case execution,
which are the dominant factors in APR efficiency. 
While each of these techniques has the potential to reduce program repair costs, they typically only affect a particular phase of the APR process.
The combination of slicing with APR has the potential to improve all phases of APR, including fault localization, patch generation, and patch validation. 

While this paper studies the impact of program reduction on a template-based APR, 
the impact on recent \emph{neural APR} approaches that build on large language models (LLMs) could even be larger. 
For example, recent work that combines various LLM-based APR approaches reported generating plausible patches for the bugs in Defects4J in 2.5 hours \emph{using ``perfect fault localization''}, i.e., using known bug locations and omitting the fault localization step~\cite{xia2023:automated}. 
Another study that does not assume perfect fault localization reported that most plausible patches are produced in 1 to 10 hours, and
half of them are generated in 4.5 hours~\cite{ye2024:iter}.
Reducing the size of the program to consider should greatly reduce these times, 
and the corresponding test-suite reduction will further reduce the time needed for patch validation.

\head{Fault localization in APR}
Fault localization is one of the key phases of the APR process because it affects not only the performance but also the correctness of generated patches~\cite{assiri2017:fault,pearson2017:evaluating,yang2020:evaluating,liu2019:you}.
Unfortunately, fault localization still suffers from accuracy issues~\cite{liu2020:efficiency,liu2019:you}.
Moreover, some classes of APR tools are more sensitive to its accuracy than others and a recent study by Liu et al. showed that
template-based APR is highly sensitive to fault localization noise~\cite{liu2020:efficiency}.

Several papers aim to improve fault localization by reducing the program spectra (i.e., the code coverage data) that need to be analyzed. 
The reduction approaches range from delta debugging~\cite{christi2018:reduce}, program slicing~\cite{zhang2007:study,alves2011:faultlocalization,li2020:more,soremekun2021:locating,soha2023:efficiency}, test generation~\cite{liu2017:improving}, to test prioritization~\cite{zhang2017:boosting}. 
In particular, Zhang et al.~\cite{zhang2007:study} evaluate the effectiveness of three variants of dynamic slicing algorithms. 
Li and Orso~\cite{li2020:more} proposed the concept of potential memory-address dependence to improve the accuracy of dynamic slicing. 
Recently, Soha~\cite{soha2023:efficiency} conducted a thorough analysis of combining various slicing and spectrum-based fault localization algorithms. 
In contrast to our work, these papers do not empirically examine the effects of test-suite and code reduction, 
nor how the fault localization improvement obtained through slicing affects the efficiency of APR.
Closer to our work, Guo et al.~\cite{guo2018:empirical} analyze how \sfs{Java} slices
impact the constraint-based APR tool Nopol~\cite{xuan2017:nopol} through its
impact on the spectrum-based fault localization used by Nopol.
They show how dynamic slicing impact on fault localization lowers \NPC and \RT.

\head{Other applications of program reduction}
Program slicing was originally developed as a debugging technique to help developers focus by reducing a program 
to exactly those statements relevant for the behavior at a point of interest~\cite{weiser1981:program, weiser1982:programmers}.
Since those early days, the abstraction capabilities of slicing have been used for various software engineering challenges, ranging from software testing, software maintenance and evolution, to finding elements for reuse~\cite{harman2001:overview, silva2012:vocabulary}.

A related idea applies dynamic Slicing to improve (spectrum-based) fault
Localization~\cite{reis2019:demystifying}.
The key intuition in this work is that highly ranked statements not covered by
failing executions are likely unrelated to the fault.
The paper shows that dynamic slicing substantially improves the rank of faulty
statements.
Our fault location results mirror these thus improving the validity of both
studies.

Of particular interest in the context of this paper is the application of program slicing to reduce the scope of what needs to be analyzed in program verification and model checking~\cite{hatcliff2001:using, chalupa2021:symbiotic}, which has been shown to be an effective way to reduce verification time~\cite{dwyer2006:evaluating, chalupa2019:evaluation}.

\section{Concluding Remarks and future work}

\noindent
Automatic program repair of large-scale programs is a difficult and
time-consuming task. 
It can take hours, and in many situations APR tools are unable to exhaustively
search the patch space.  
Prior work showed that the most challenging parts of the repair process are
finding faulty statements and verifying the generated patch's correctness.  
Instead of examining an entire program, this work aims to focus the repair process
on a relatively small portion of the program in the form of a program slice.

The experiments show that slicing \emph{can} significantly improve the performance of
template-based APR, but not in all cases. 
The benefit is primarily through test-suite size reduction.
Like most research problems, this investigation raises as many question as it answers.
Future work will consider addressing the limitations we encountered, such as
how to better exploit the reduced program in fault localization.
We also intend to consider other classes of APR tools, mainly neural and
semantic-based APR tools. 

Finally, we hope to develop a better understanding of TBar's inability to patch
\PS and the poor performance of GZoltar when applies to \PS and \TS.
One possibility here is that as libraries, \sfs{Math} and \sfs{Lang} might
prove harder to fix because they include less cohesive code, which makes it
more likely that the code violates the plastic surgery
hypothesis~\cite{barr2014:plastic}.
Non-library slices are expected to be larger and more cohesive making them
more likely to include reusable code.  
Thus we plan experiments that considers projects with a range of cohesiveness
to see if there is any correlation with TBar's and GZoltar's performance. 

\section*{Acknowledgements}

\noindent
The research presented in this paper was financially supported by the Research Council of Norway through the secureIT project (grant \#288787). 

The authors wish to thank Lulu Wang for providing JavaSlicer and assisting in getting it to run for our experiments.

{\balance
\printbibliography 
}
\end{document}